\documentclass[preprint,12pt]{elsarticle}

 \usepackage{graphicx}

\usepackage{amssymb}

\newcommand{\scl}{0.63}
\newcommand{\C}{{}^{12}\mathrm{C}}
\newcommand{\Cn}{{}^{13}\mathrm{C}}
\newcommand{\A}[2]{{}^{#1}\mathrm{#2}}
\newcommand{\Ox}{{}^{16}\mathrm{O}}
\newcommand{\On}{{}^{17}\mathrm{O}}

\journal{Nuclear Physics A}

\begin{document}

\begin{frontmatter}

\title{Spin observables in three-body direct nuclear reactions}


\author{A. Deltuva}
\ead{deltuva@cii.fc.ul.pt}
\address{Centro de F\'{\i}sica Nuclear da Universidade de Lisboa,
  P-1649-003 Lisboa, Portugal}

\begin{abstract}
Direct nuclear reactions $\vec{d}+A$ and $\vec{p}+(An)$
are described in the framework of three-body Faddeev-type equations.
Differential cross section and analyzing powers are calculated using
several optical potential models and compared with the experimental data.
Quite satisfactory agreement is found except for few systematic 
discrepancies.
\end{abstract}

\begin{keyword}
three-body scattering \sep Faddeev equations \sep analyzing power

\PACS 24.10.-i \sep 21.45.+v \sep 24.70.+s \sep  25.55.Ci 
\end{keyword}

\end{frontmatter}


\section{Introduction}

Direct nuclear reactions, dominated by three-body degrees of freedom,
provide an important test for the models of nuclear dynamics.
In the past, the reactions like the deuteron $(d)$ scattering from a stable
nucleus $(A)$ were described using the approximate 
continuum discretized coupled channels (CDCC) method \cite{austern:87}.
With few exceptions the spin degrees of freedom have been usually neglected.
In such a case one can only calculate the unpolarized cross sections
which are mostly sensitive to the central part of the nucleon-nucleus
optical potential. However, there exist  experimental data
for polarization observables that depend as well on the spin-orbit part
of the optical potential. The aim of the present work is to study
the spin observables in $\vec{d}+A$ and $\vec{p}+(An)$ reactions
in the framework of exact three-body Faddeev/Alt, 
Grassberger, and Sandhas (AGS) equations \cite{faddeev:60a,alt:67a}
such that all discrepancies with the experimental data can be attributed
solely to the shortcomings of the used optical potential models.
A part of the reactions to be considered here have already been studied
by the CDCC method \cite{iseri:88}.
Thus, the comparison with the results of Ref.~\cite{iseri:88}
may allow us to draw some conclusions on the reliability of CDCC for
calculating spin observables, in addition to the cross section benchmark
\cite{deltuva:07d}.

The theoretical framework is shortly recalled in Sec.~\ref{sec:th},
the results are presented in Sec.~\ref{sec:res}, and the summary is given
in Sec.~\ref{sec:sum}.

\section{Faddeev/AGS equations and dynamic input}
\label{sec:th}

We describe  $\vec{d}+A$ and $\vec{p}+(An)$ type reactions using three-body 
$(p,n,A)$ model. An exact treatment of the quantum three-body scattering problem
is provided by both Faddeev \cite{faddeev:60a} and AGS equations \cite{alt:67a}
that are equivalent to the Schr\"odinger equation
but are more suitable for the numerical solution
due to the connectedness of the kernel. The Faddeev equations are formulated
for the components of the wave function while the AGS equations, 
\begin{equation}  \label{eq:Uba}
U_{\beta \alpha}  = \bar{\delta}_{\beta\alpha} \, G^{-1}_{0}  +
\sum_{\sigma=1}^3   \bar{\delta}_{\beta \sigma} \, T_{\sigma} 
\, G_{0} U_{\sigma \alpha},
\end{equation}
are a system of coupled integral equations for the transition operators 
$U_{\beta \alpha}$ whose on-shell matrix elements 
$\langle\psi_{\beta}|U_{\beta \alpha}|\psi_{\alpha}\rangle$
are scattering amplitudes and therefore lead directly to the  observables.
In Eq.~(\ref{eq:Uba}) $ \bar{\delta}_{\beta\alpha} = 1 - \delta_{\beta\alpha}$,
$G_0 = (E+i0-H_0)^{-1}$ is the free resolvent,
and $T_{\sigma} = v_{\sigma} + v_{\sigma} G_0 T_{\sigma}$ is the two-particle
transition matrix, $E$ being the available three-particle energy in the 
center of mass (c.m.) system, $H_0$ the free Hamiltonian, and 
$v_{\sigma}$ the potential for the pair $\sigma$ in odd-man-out notation.
The channel states $|\psi_{\sigma}\rangle$ for  $\sigma = 1,2,3$ are the
eigenstates of the corresponding channel Hamiltonian $H_\sigma = H_0 + v_\sigma$
with the energy eigenvalue $E$; thus, $|\psi_{\sigma}\rangle$ is a product of
the bound state wave function for pair $\sigma$ and a plane wave 
with fixed on-shell momentum
corresponding to the relative motion of particle $\sigma$ and pair $\sigma$
in the initial or final state. Observables of elastic scattering
are calculated from the matrix elements with
$\beta = \alpha$ while $\beta \neq \alpha$ corresponds to transfer reactions. 

We solve the AGS equations using momentum-space partial-wave basis
where they become a system of integral equations
with two continuous variables, the values of Jacobi momenta. 
The employed numerical techniques are described in great detail in 
Refs.~\cite{chmielewski:03a,deltuva:03a,deltuva:05a}.

The  AGS equations are applicable only to short-range potentials $v_{\sigma}$. 
Nevertheless, the long-range Coulomb force between charged particles can
be included in this framework using the method of screening and renormalization
\cite{taylor:74a,alt:80a,deltuva:08c} which enables 
to calculate the Coulomb-distorted short-range part of the
transition amplitude by solving the AGS equations with nuclear plus 
screened Coulomb potential; the convergence of the results with the 
screening radius has to be established.
The method has been successfully applied to proton-deuteron
\cite{deltuva:05a,deltuva:05c}
and $\alpha$-deuteron \cite{deltuva:06b} elastic scattering and breakup, and to
three-body nuclear reactions involving deuterons or one-neutron halo nuclei
\cite{deltuva:07d}.

The dynamic input to the AGS equations are the potentials $v_{\sigma}$
for the three pairs of particles.
As the $np$ interaction we take realistic
CD Bonn potential \cite{machleidt:01a}, in contrast to the usual CDCC
calculations where a simple Gaussian  $np$ potential is used.
For the nucleon-nucleus $(NA)$ interaction, in order to study the model 
dependence, we use several different  optical potentials, namely, 
those by Watson et al.~\cite{watson}, Menet et al.~\cite{menet},
Becchetti and Greenlees \cite{becchetti}, and Koning and Delaroche 
\cite{koning}; the corresponding predictions in the following will be
abbreviated by W, M, BG, and KD, respectively. 
Each of these potentials is fitted to the $NA$ data in a limited mass 
and energy range; those limitations are respected in the present 
calculations. The energy-dependent parameters of the potentials are taken
at a half deuteron lab energy in the $\vec{d}+A$ reactions
and at the proton lab energy for $pA$ in the $\vec{p}+(An)$ reactions. 
In the latter case the $nA$ potential
is real and supports a number of bound states corresponding
to the ground and excited single-particle states of the $(An)$ nucleus
while all Pauli forbidden states are removed;
the potential parameters and the resulting binding energies are given
in Ref.~\cite{deltuva:09a} for  $\Cn$ and $\On$ nuclei.
The interaction within $np$, $nA$, and $pA$ pairs is included in partial
waves with pair orbital angular momentum $L \leq 3$, 10, and 20, respectively,
and the total angular momentum is $J \leq 45$; depending on the reaction
some of these quantum numbers cutoffs can be safely chosen significantly lower,
leading, nevertheless, to well converged results.
The $pA$ channel is more demanding than the $nA$ channel due to the screened
Coulomb force, where the screening radius $R \approx 8$ to 10 fm for the
short-range part of the scattering amplitude is sufficient for the
convergence.

\section{Results} \label{sec:res}

The experimental data for the spin observables are scarcer than for
the spin-averaged cross sections. Nevertheless, a complete set of
deuteron analyzing powers over a wide mass and angular range 
is presented in Ref.~\cite{matsuoka:86} for $E_d = 56$ MeV; other measurements
exist as well but cover only rather narrow angular or mass range.
We therefore concentrate first on 56 MeV polarized deuteron scattering
from $\C$, $\Ox$, $\A{28}{Si}$, $\A{40}{Ca}$, and $\A{58}{Ni}$ nuclei.
The results are presented in Figs.~\ref{fig:dCO} and \ref{fig:dSiCaNi}.

\renewcommand{\scl}{0.55}
\begin{figure}[!]
\begin{center}
\includegraphics[scale=\scl]{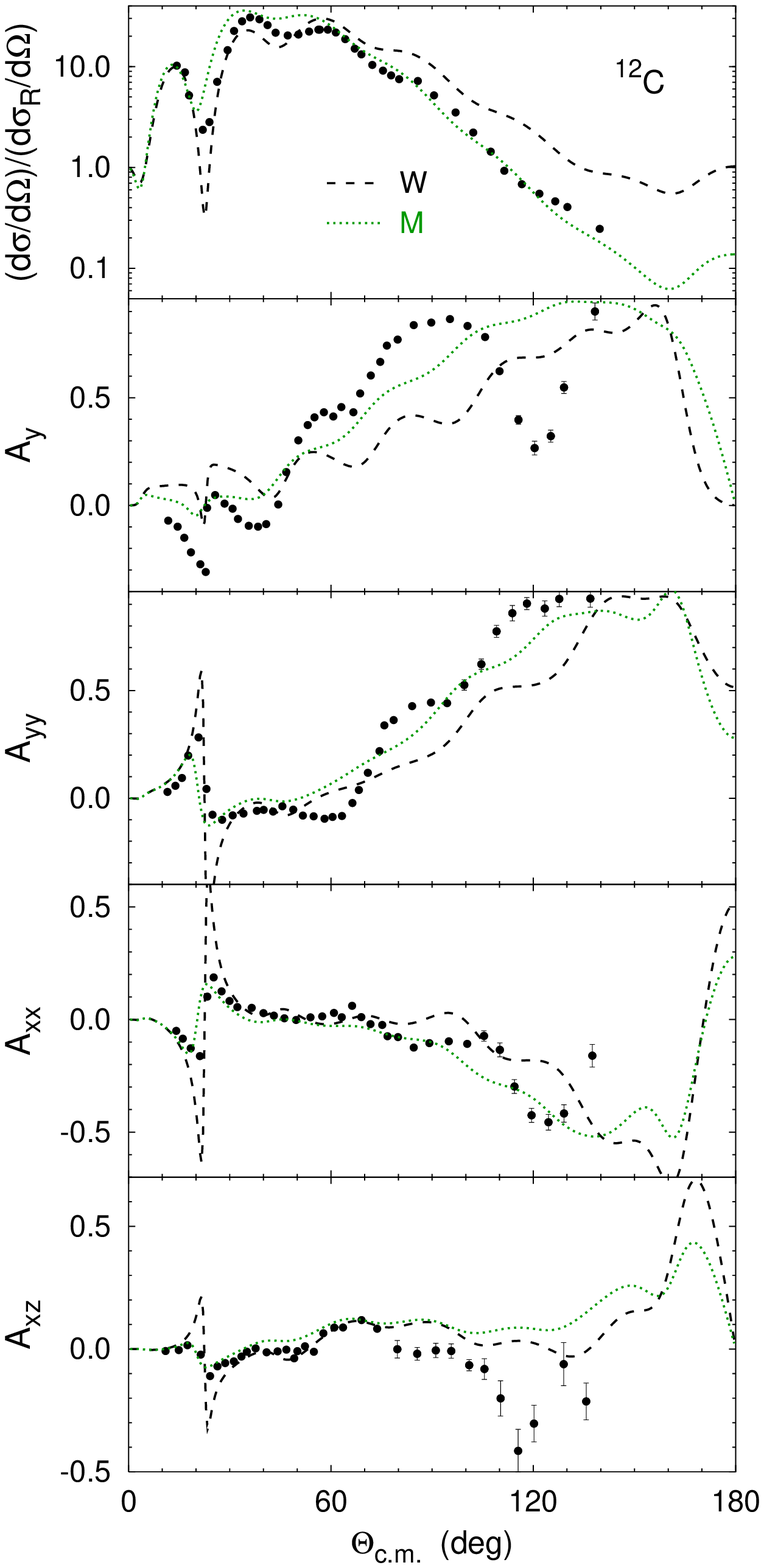} \hspace{-5mm}
\includegraphics[scale=\scl]{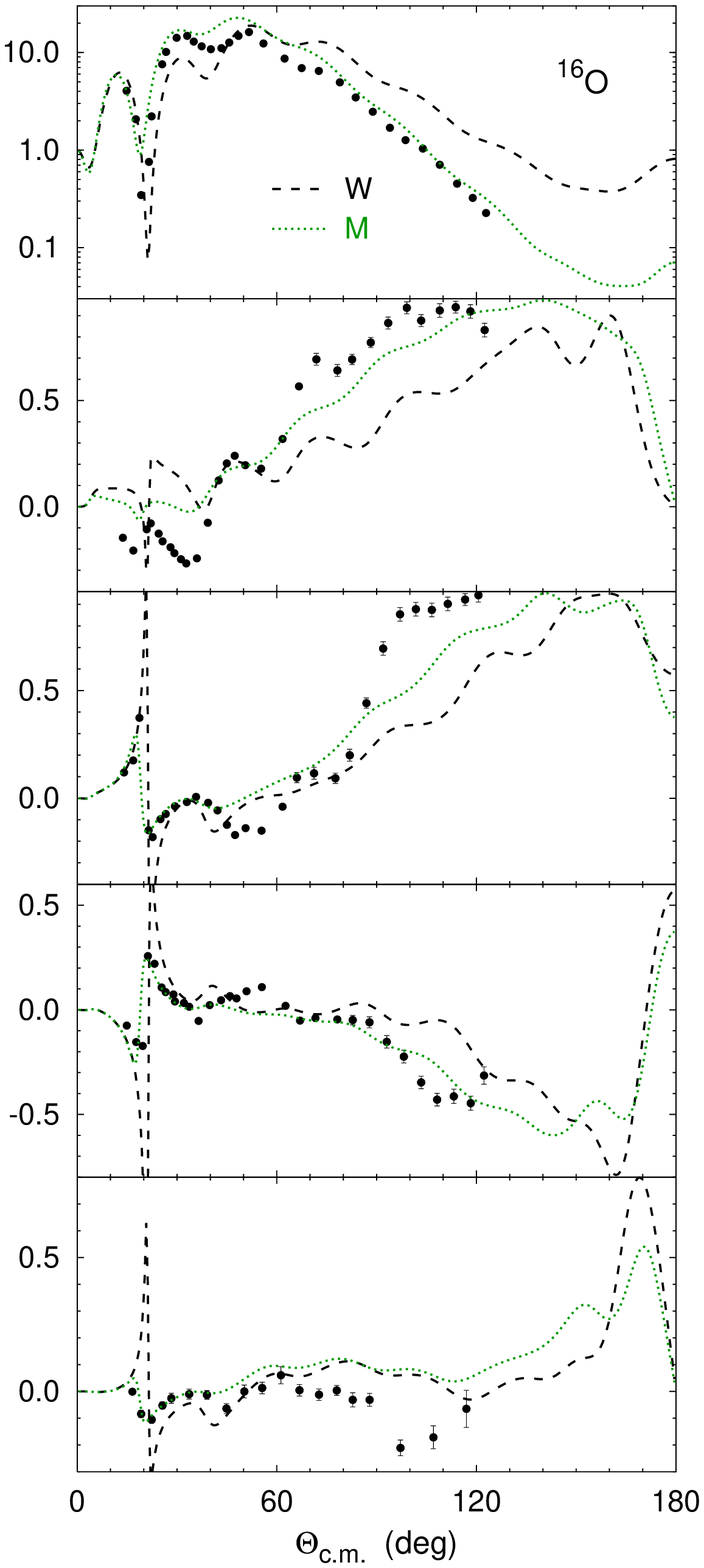}
\end{center}
\caption{\label{fig:dCO}
Differential cross section divided by Rutherford cross section
and deuteron analyzing powers for  deuteron elastic scattering
from  $\C$ and $\Ox$ nuclei at $E_d = 56$ MeV
as functions of the c.m. scattering angle.
Results for $NA$ potentials W and M are given by dashed and dotted curves,
respectively. The experimental data are from Ref.~\cite{matsuoka:86}.}
\end{figure}

\renewcommand{\scl}{0.55}
\begin{figure}[!]
\begin{center}
\includegraphics[scale=\scl]{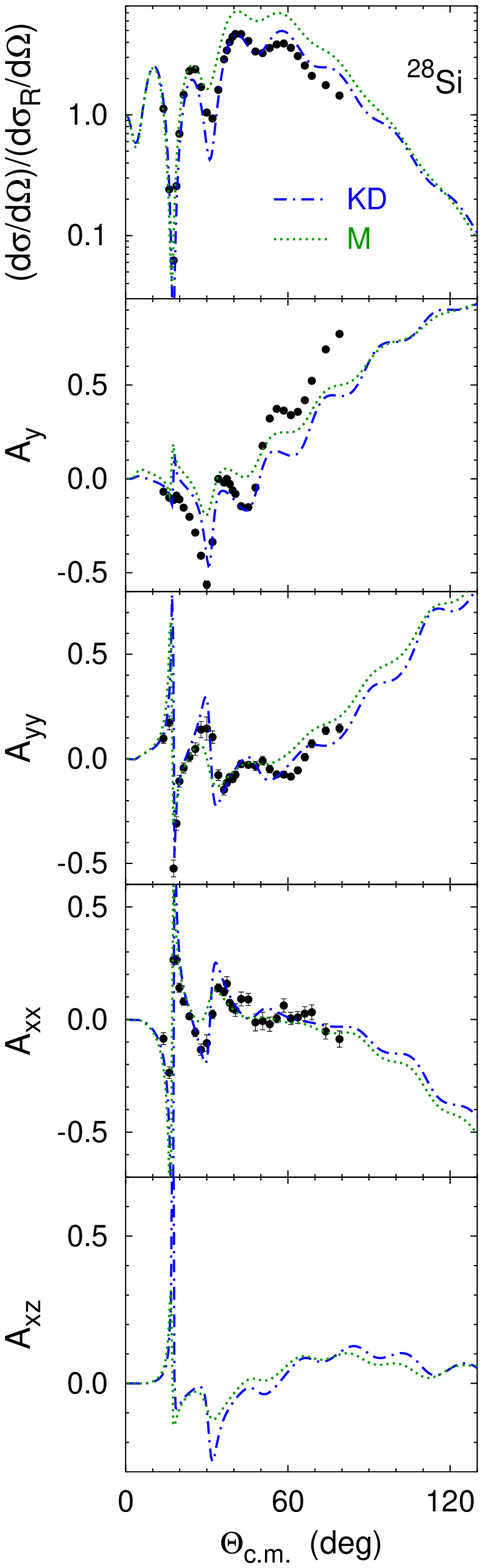} \hspace{-5mm}
\includegraphics[scale=\scl]{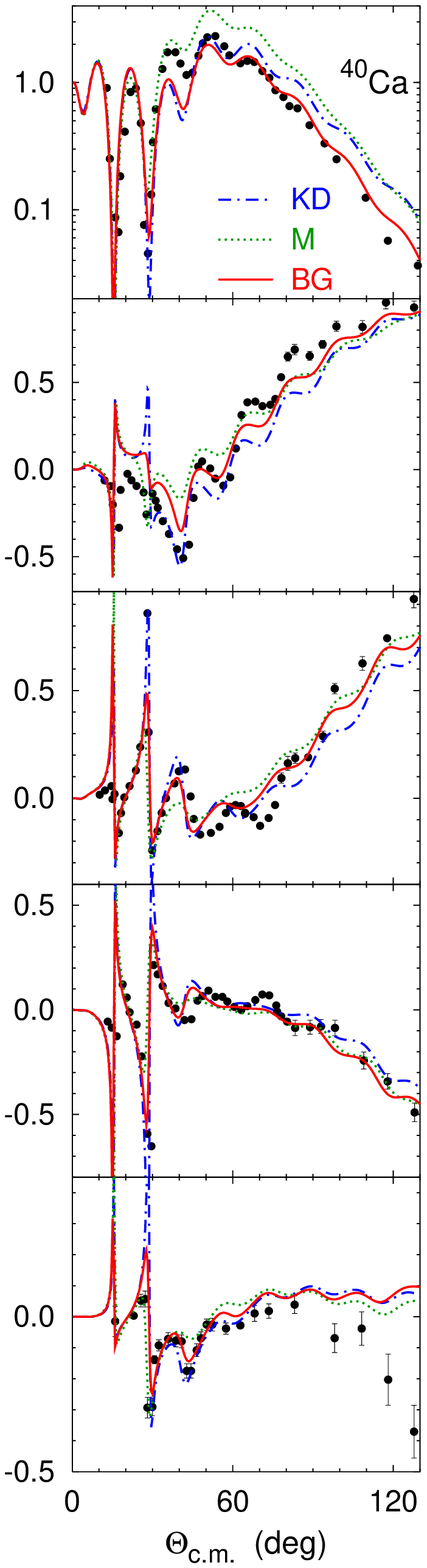} \hspace{-5mm}
\includegraphics[scale=\scl]{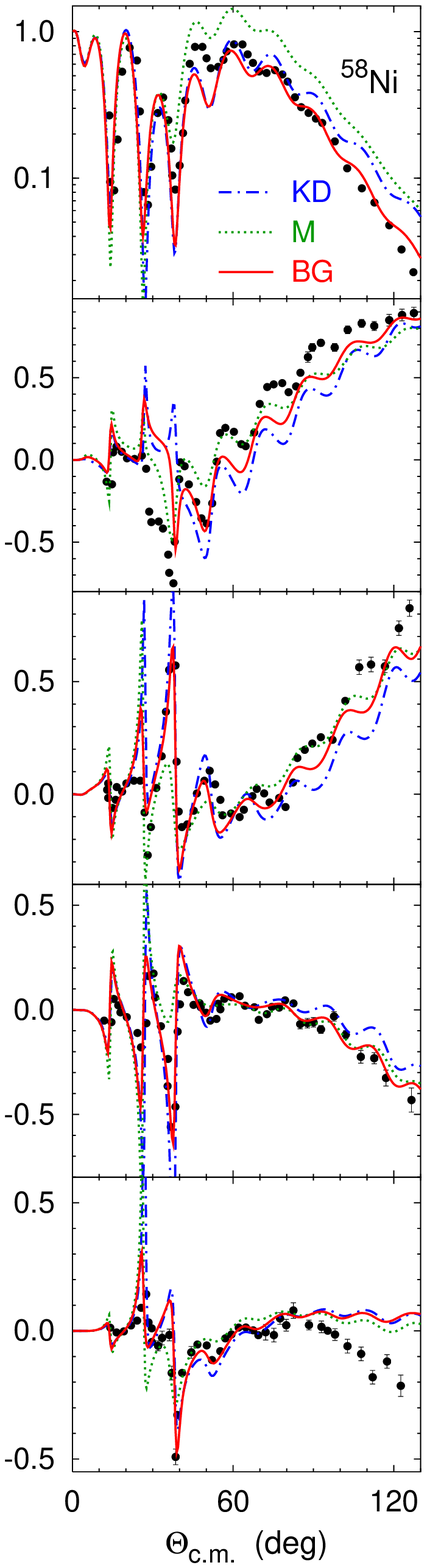}
\end{center}
\caption{\label{fig:dSiCaNi}
Differential cross section divided by Rutherford cross section
and deuteron analyzing powers for  deuteron elastic scattering
from $\A{28}{Si}$, $\A{40}{Ca}$, and $\A{58}{Ni}$ nuclei at $E_d = 56$ MeV.
Results obtained with $NA$ potentials KD, M, and BG are given by 
dashed-dotted, dotted, and solid curves, respectively. 
The experimental data are from Ref.~\cite{hatanaka:80} for $\A{28}{Si}$
and from Ref.~\cite{matsuoka:86} for other nuclei.}
\end{figure}

As shown in Fig.~\ref{fig:dCO},
in the case of $\C$ and $\Ox$ nuclei the predictions using the potential M 
describe the experimental data better than those of W, especially at 
larger scattering angles. There are significant differences also 
around $\Theta_{c.m.} = 22$ deg where W predicts much deeper minimum
in the cross section  and thereby too sharp peaks in the analyzing powers;
in contrast, the corresponding peaks obtained with M  for  $\C$
are not sharp enough. Both potentials fail to reproduce the vector
analyzing power $A_y$ data at small angles  $\Theta_{c.m.} < 40$ deg;
the reason for this failure, at least to some extent,
 may be the inability to describe  $A_y$ at
small angles in $\vec{p} + A$ elastic scattering as can be seen in
Ref.~\cite{watson}. The $A_y$ and $A_{yy}$ data at intermediate angles are 
slightly underpredicted, and, in addition, $A_y$ for $\C$ shows a clear 
minimum around $\Theta_{c.m.} = 120$ deg that is not present in the
theoretical results. Furthermore, calculated $A_{xz}$ for both $\C$ and 
$\Ox$ nuclei at larger angles deviates quite significantly from the data.
Otherwise the qualitative description of the data by the potential M
is quite satisfactory. The comparison with the $\vec{d}+\C$ data
from Ref.~\cite{kato:85} at $E_d = 35$ to 70 MeV that are limited to narrow
angular range from 35 to 80 deg (not shown here) 
brings essentially the same conclusions.

However, the potential M fails in the case of heavier nuclei $\A{28}{Si}$,
$\A{40}{Ca}$, and $\A{58}{Ni}$ as shown
in  Fig.~\ref{fig:dSiCaNi}. It clearly overpredicts the differential cross
section at $\Theta_{c.m.} > 40$ deg while the analyzing power data are 
reproduced with similar quality as by other potentials.
Quite surprisingly,  at larger angles $\Theta_{c.m.} > 60$ deg
the old global potential BG describes deuteron scattering
from $\A{40}{Ca}$ and $\A{58}{Ni}$ nuclei better than the new and precise 
KD potential which has many more parameters fitted individually 
to the $nA$ and $pA$ data of the considered nuclei.
At smaller angles all potentials account for the data with
comparable quality except for $A_y$ where KD shows an additional sharp maximum
instead of a minimum. The most serious discrepancies between theory and 
data take place in small-angle $A_y$ 
and in large-angle $A_{xz}$. While the latter one is
similar as in the case of  $\C$ and $\Ox$ nuclei, the former one
is somehow different: the data are reproduced at very small angles,
but then there is a narrow angular interval around  $\Theta_{c.m.} = 30$ deg
where $A_y$ is strongly overpredicted. Furthermore, in contrast to
 $\C$ and $\Ox$, $A_y$ in proton 
scattering from heavier nuclei is well described by the employed
potentials. Thus, in this case the $A_y$ discrepancy even partially cannot be
explained by a poor $NA$ data fit and must be due to another, presently 
unknown, reason.
In addition, $A_y$ and $A_{yy}$ at larger angles are slightly underpredicted
but the oscillating behaviour of the observables is reproduced rather well.
Otherwise the qualitative description of the data, especially 
by the potential BG, is quite satisfactory.

Even larger discrepancy in the vector analyzing power $A_y$ or, equivalently, 
$iT_{11} = A_y \sqrt{3}/2$, is present at lower energies
as shown in Fig.~\ref{fig:dCSi29} for the elastic deuteron scattering 
from $\C$ and $\A{28}{Si}$ nuclei at $E_d = 29.5$ MeV.
Compared to Figs.~\ref{fig:dCO} and \ref{fig:dSiCaNi}, the discrepancy
takes place at larger scattering angles, 30 to 55 deg and 45 to 70 deg
for $\C$ and $\A{28}{Si}$, respectively.
Outside those regions the calculations  account for the data quite
satisfactorily, at least in the case of $\A{28}{Si}$.
The agreement for the differential cross section is rather similar to the 
one seen at  $E_d = 56$ MeV.

\renewcommand{\scl}{0.54}
\begin{figure}[t]
\begin{center}
\includegraphics[scale=\scl]{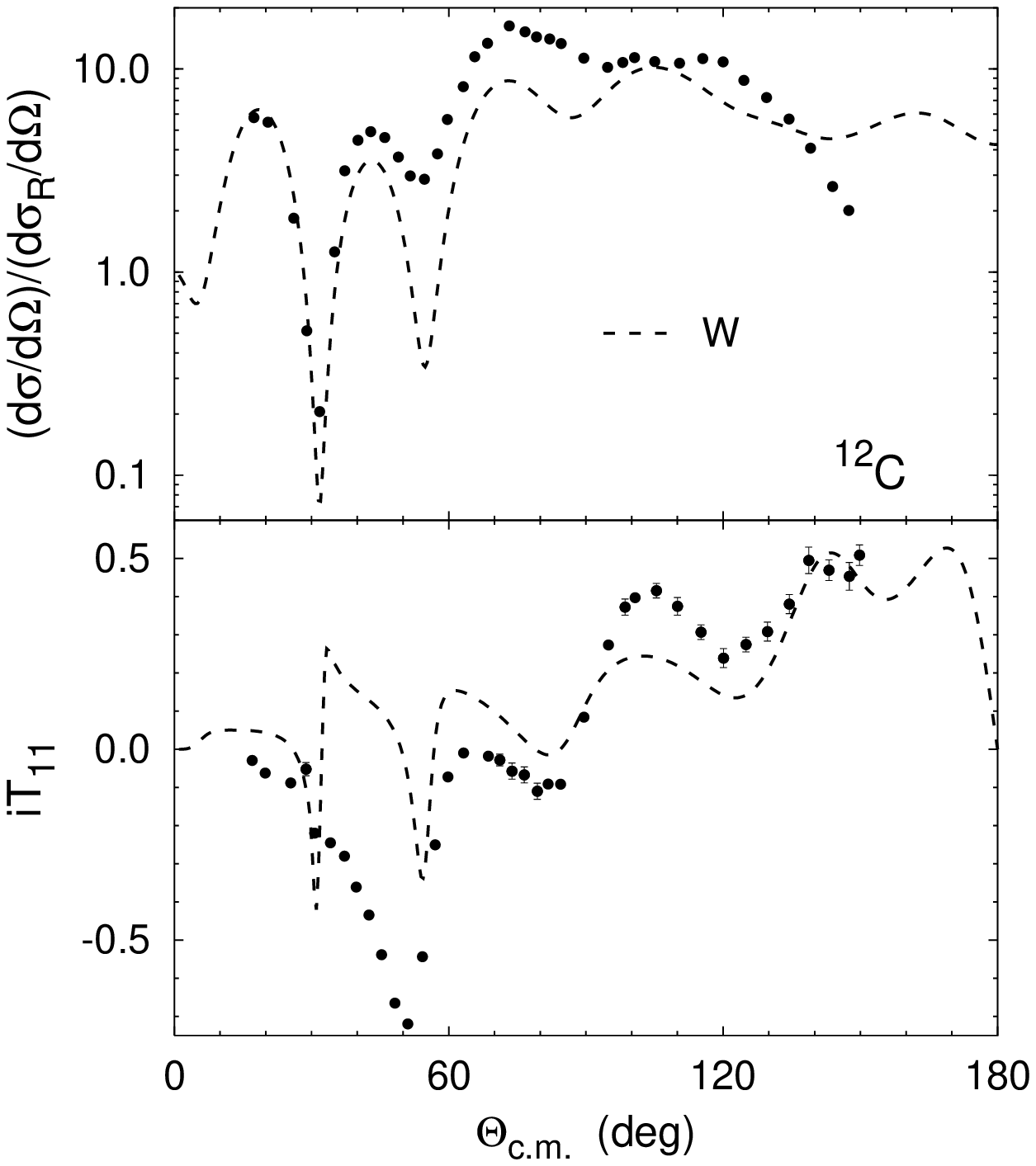} \hspace{-5mm}
\includegraphics[scale=\scl]{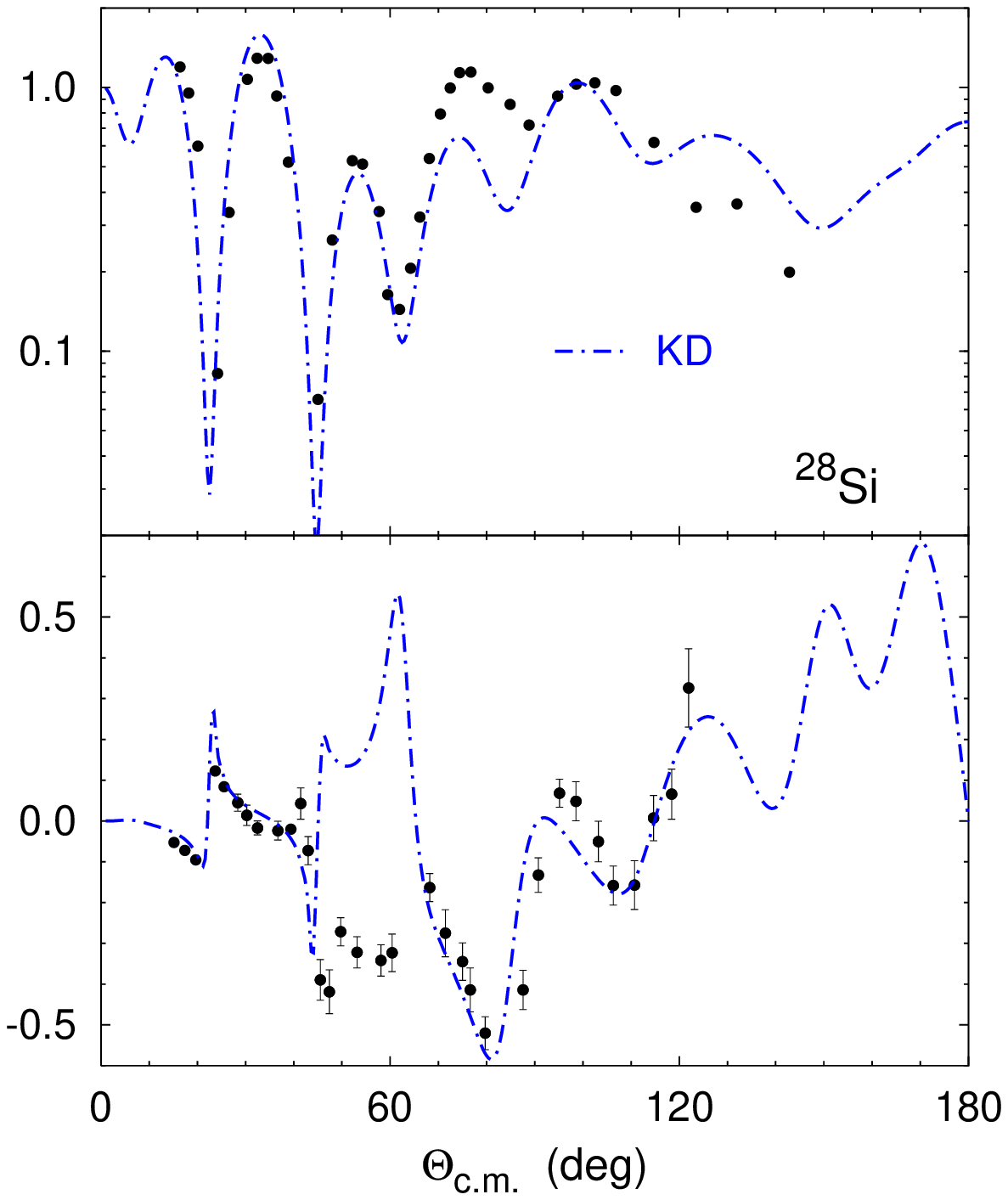}
\end{center}
\caption{\label{fig:dCSi29}
Differential cross section divided by Rutherford cross section
and deuteron vector analyzing power $iT_{11}$ for deuteron 
elastic scattering from $\C$ and  $\A{28}{Si}$ nuclei at $E_d = 29.5$ MeV
calculated using the potentials W and KD, respectively.
The experimental data are from Ref.~\cite{perrin:77}.}
\end{figure}

Deuteron scattering from $\Ox$, $\A{40}{Ca}$, and $\A{58}{Ni}$ nuclei at 
$E_d = 56$ MeV was calculated in Ref.~\cite{iseri:88} in the CDCC framework
using M and BG optical potentials. However, 
the $np$ interaction in Ref.~\cite{iseri:88} and in the present work is not
 exactly the same.  In Ref.~\cite{iseri:88} the
deuteron bound state was calculated with a realistic $np$ potential
but the $np$ continuum was described by simple Gaussian potential acting
in the triplet S and D waves only. Such a mixed potential cannot be included
in consistent Faddeev-type calculations such as of the present work. 
Nevertheless, since the considered observables are rather insensitive to
the $np$ potential, a comparison of the present results
and those of  Ref.~\cite{iseri:88} is meaningful. 
The agreement is good, there are only rather small differences,
e.g., our $A_y$ values for $\Theta_{c.m.} \geq 60$ deg are higher by about
0.05 -  0.10 coming closer to the data. Thus, in addition to the cross
section benchmark \cite{deltuva:07d} one can conclude that
CDCC is quite reliable also in calculating the spin observables in
deuteron-nucleus elastic scattering.

The $\vec{p}+(An)$ elastic scattering is a less interesting case since 
the differential cross section and the proton analyzing power are 
quite strongly correlated with the corresponding observables
in $\vec{p}+A$ elastic scattering to which the $pA$ potential is fitted.
The data for other observables like $(An)$ analyzing power and 
spin correlation coefficients does not exist. Furthermore,
the only available polarization data refer to  proton-$\Cn$ scattering.
Examples at proton lab energy $E_p =17.5$ and 35 MeV  are shown
in Fig.~\ref{fig:pC} together with the  differential cross section. 
The calculations describe the experimental data with the same 
quality as for the corresponding proton-$\C$ observables and even
show similar deviations \cite{watson}, e.g., underestimation of the large angle
 cross section  at 17.5 MeV and overestimation of the small angle $A_y$.
However, in contrast to the deuteron-$\C$ scattering, it seems that 
the potential W describes the data slightly better than M, except
for forward angles.

\renewcommand{\scl}{0.54}
\begin{figure}[!]
\begin{center}
\includegraphics[scale=\scl]{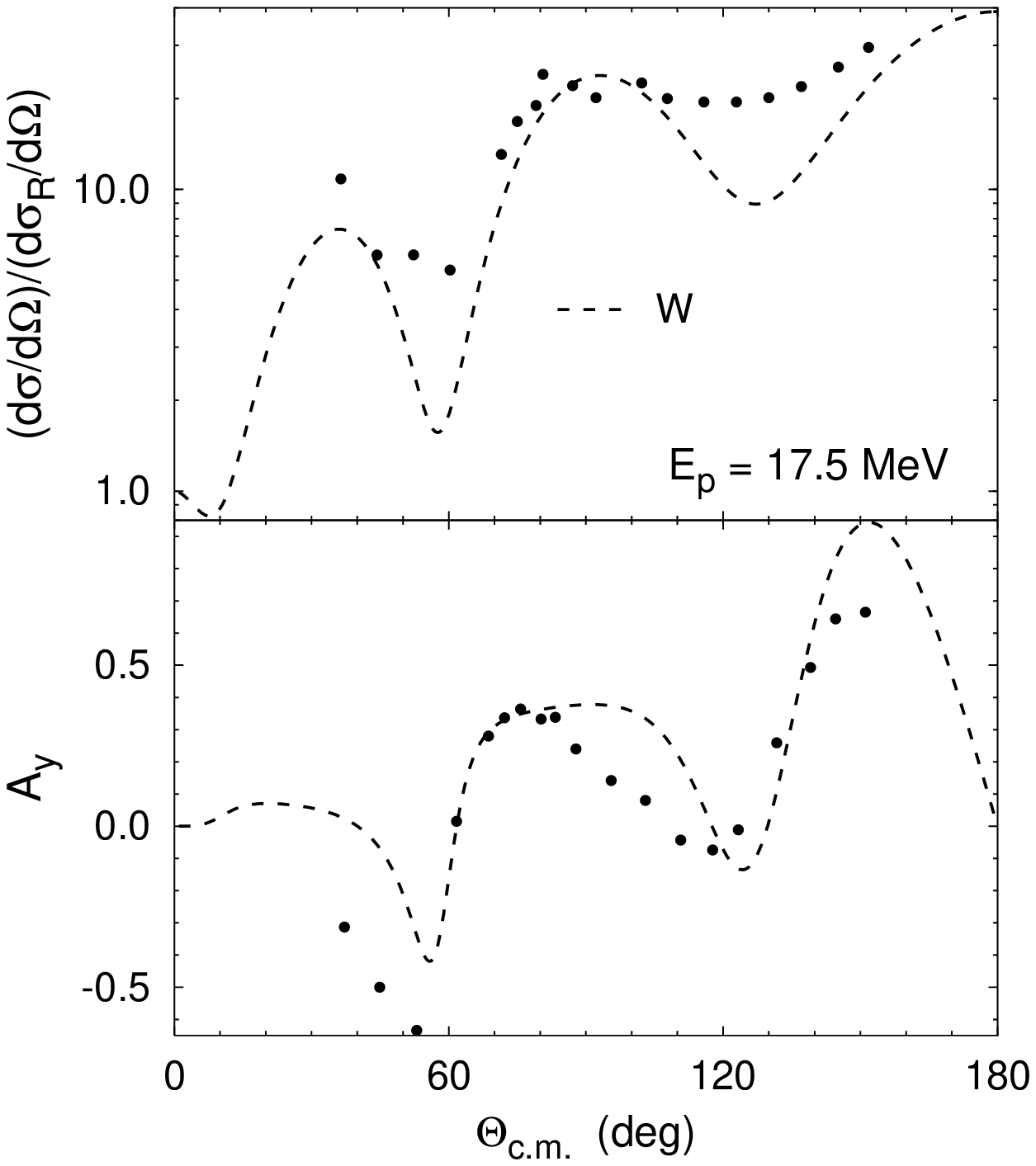} \hspace{-5mm}
\includegraphics[scale=\scl]{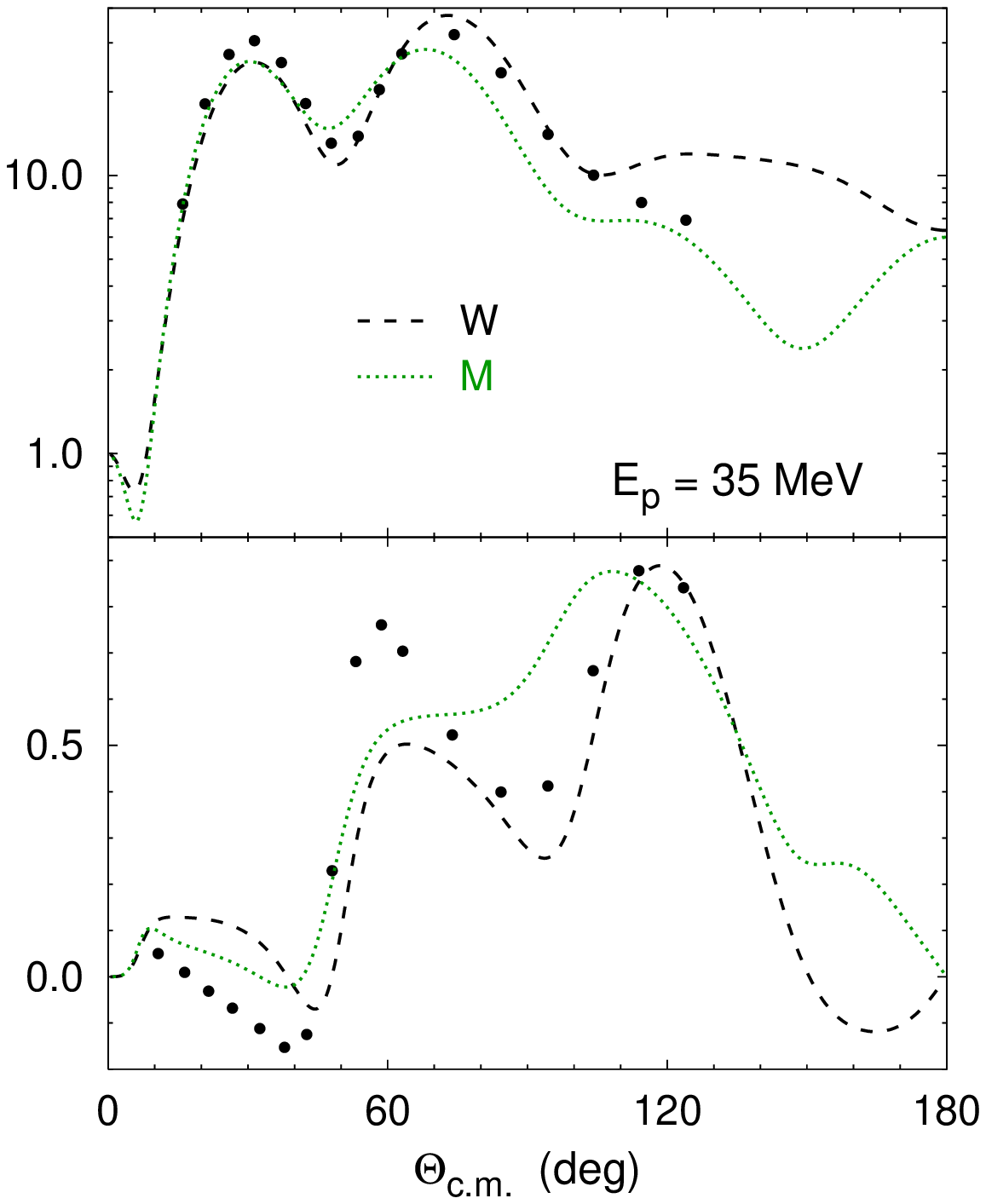}
\end{center}
\caption{\label{fig:pC}
Differential cross section divided by Rutherford cross section
and proton analyzing power for proton-$\Cn$ elastic scattering
at $E_p = 17.5$ and 35 MeV.
Curves as in Fig.~\ref{fig:dCO}.
The experimental data are from Refs.~\cite{pC17}
and \cite{pC35} at 17.5 and 35 MeV, respectively.}
\end{figure}

\renewcommand{\scl}{0.56}
\begin{figure}[!]
\begin{center}
\includegraphics[scale=\scl]{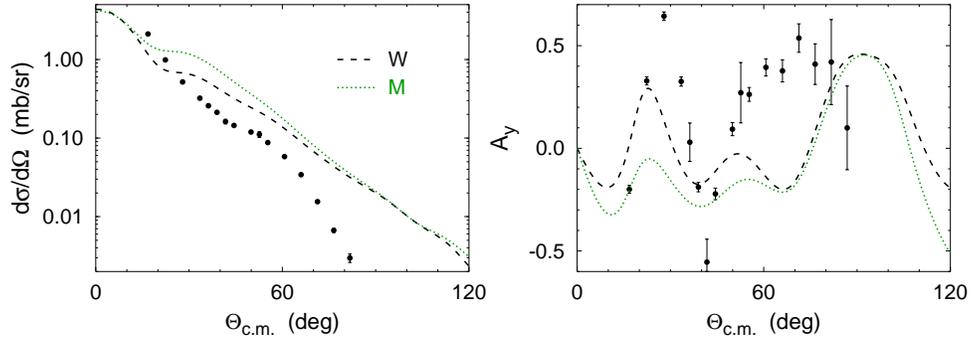}
\end{center}
\caption{\label{fig:pC65d}
Differential cross section and proton analyzing power for 
$\vec{p} + \Cn \to d+ \C$ reaction at $E_p = 65$ MeV.
Curves as in Fig.~\ref{fig:dCO}.
The experimental data are from Ref.~\cite{pC65d}.}
\end{figure}

In the Faddeev/AGS framework the elastic scattering and
the transfer reaction $\vec{p} + (An) \to d + A$ are calculated 
simultaneously. Unfortunately, the data only exists for
$\vec{p} + \Cn \to d + \C$ transfer at $E_p = 65$ MeV.
Although this is 
slightly above the upper validity limit of the employed optical 
potentials W and M and both of them do not reproduce accurately
the $pA$ data at this energy, we show the results in
Fig.~\ref{fig:pC65d}. The description of the data is poor for
both differential cross section and proton analyzing power,
and is slightly worse when using the potential M.
Note that a similar failure in the transfer cross section is 
observed in the $d+\Ox \to p+\On$ reaction 
at $E_d = 63.2$ MeV \cite{deltuva:09a}.

\section{Summary} \label{sec:sum}

We performed Faddeev-type calculations of three-body direct nuclear
reactions in $(p,n,A)$ model.
 The framework is the integral momentum-space AGS equations;
the Coulomb interaction between the charged particles is included using
the method of screening and renormalization; well converged results
are obtained. Thus, all discrepancies with the experimental data can be 
attributed to the shortcomings of the used optical potentials
or even to the inadequacy of the three-body model.
A realistic $np$ potential and several parametrizations of the 
nucleon-nucleus optical potentials were used, all including spin-orbit
interaction. Differential cross section and analyzing powers were
calculated for the deuteron elastic scattering from
$\C$, $\Ox$, $\A{28}{Si}$, $\A{40}{Ca}$, and $\A{58}{Ni}$ nuclei,
and for $\vec{p} + \Cn$ elastic scattering and transfer to $d + \C$.
The description of the experimental data is mostly quite satisfactory,
at least by some of the optical potential models, and is of similar
quality as in the nucleon-nucleus elastic scattering to which 
the optical potentials are fitted.
Systematic discrepancies are found in elastic proton and deuteron vector 
analyzing power $A_y$ at small angles, and in the deuteron tensor analyzing 
power $A_{xz}$ at large angles. The former one, in the case of light
nuclei $\C$ and $\Ox$,  may be related to a
similar problem in the nucleon-nucleus $A_y$.
The calculations also fail in accounting for the available
$\vec{p} + \Cn \to d + \C$ transfer data.
A satisfactory qualitative agreement with previous CDCC results for
$\vec{d}+A$ elastic observables is found indicating
the reliability of the CDCC method for this type of reactions.

\section*{Acknowledgements}
The author thanks A.~C.~Fonseca for comments on the manuscript.
The work is supported by the Funda\c{c}\~{a}o para a Ci\^{e}ncia
e a Tecnologia (FCT) grant SFRH/BPD/34628/2007.




\end{document}